\documentclass{edbk} 
  \usepackage{psfig}

\kluwerbib

\begin{document}

\long\def\jumpover#1{{}}
\def\ni{\noindent}
\def\th{\thinspace}
\def\ngth{\negthinspace}
\def\ngthh{\negthinspace\negthinspace\negthinspace}
\def\sss#1{{{\scriptscriptstyle #1}}}
\def\ah{{\scriptscriptstyle 1/2}}
\def\tq{{\scriptstyle 3/4}}
\def\Teff{{$T_{ef\!f} $}}
\def\Mo{{M$_\odot $}}
\def\Lo{{L$_\odot $}}
\def\eg{{\it e.g.~}}
\def\cf{{\it cf.~}}
\def\ie{{\it i.e.~}}
\def\viz{{\it viz.~}}
\def\vs{{\it vs.~}}
\def\etal{{\it et al.~}}
\def\apriori{{\it a priori~}}
\def\ff  {{\em ff.~}}
\def\apj{{ApJ~}}
\def\ub#1{{\underbar {#1}}}

\booktitle[Nonlinear Stellar Pulsation]
{Nonlinear Stellar Pulsation}




\articletitle{Double-Mode Stellar Pulsations}


\author{Zolt\'an Koll\'ath}
\affil{Konkoly Observatory\\
P.O. Box 67, H-1525 Budapest, Hungary}
\email{kollath@konkoly.hu}

\author{J. Robert Buchler}
\affil{Physics Department, University of Florida\\
Gainesville FL32611, USA}
\email{buchler@physics.ufl.edu}

\begin{abstract} 
The status of the hydrodynamical modelling of nonlinear multi-mode stellar
pulsations is discussed.  The hydrodynamical modelling of steady double-mode
(DM) pulsations has been a long-standing quest that is finally being concluded.
Recent progress has been made thanks to the introduction of turbulent
convection in the numerical hydrodynamical codes which provide detailed results
for individual models.  An overview of the modal selection problem in the HR
diagram can be obtained in the form of bifurcation diagrams with the help of
simple nonresonant amplitude equations that capture the DM phenomenon.
\end{abstract}


\vskip-16cm
{\large \noindent{\bf NONLINEAR STELLAR PULSATION}\\
\noindent{\bf Editors: M. Takeuti \& D. Sasselov}\\
\noindent{\bf in Astrophysics and Space Science Library (ASSL)}

\vskip 15cm

\section{Introduction}

Self-excited multi-mode pulsations are quite common among the less evolved
luminous stars, such as the delta Scuti stars, but they mostly involve
nonradial modes of oscillation.  In contrast, the classical variable stars,
\viz the Cepheids and RR~Lyrae stars, are believed to be radially pulsating,
and two independent frequencies are typically identified in the Fourier
spectrum.  These pulsating stars are referred to as double-mode even though, in
principle, there could be more than two modes involved with locked frequencies.
(We should add that there is evidence that suggests the presence for nonradial
modes as well (Kov\'acs \etal, Moskalik, Olech \etal in Szabados \& Kurtz 2000)
.

In the Galaxy, the Beat Cepheids, also called double-mode (DM) Cepheids are
relatively rare.  Only about a good dozen are known.  On the other hand, the
recent observations of the Magellanic Clouds (Beaulieu \etal 1995, 1997, Welch
\etal 1995, Udalski \etal 1999) have shown that in these galaxies Beat Cepheids
are quite common.  Beat Cepheid pulsations occur either in the
fundamental/first overtone (F/O$_1$) modes, or in the first/second overtone
(O$_1$/O$_2$) modes.  In the Galaxy there is only one O$_1$/O$_2$ known DM
pulsator (CO~Aur), but in the SMC Udalski \etal 1999 claim 70 candidates out of
93 DM Cepheids.

The occurrence of DM  RR~Lyrae stars (RRd stars), depends on the type
of cluster.  For example, not a single RRd star is observed in $\omega$
Centauri (\cf Kov\'acs in this Volume) while in M68, 9 RRd stars are
known out of 37 RR Lyrae stars.

The DM pulsators are important for pulsation theory in that they impose very
stringent requirements on the numerical modelling.  Double periodicity and
indeed the overall view of the modal selection picture provide many more
observational constraints than their single-mode siblings.  

Despite the frequent occurrence of DM pulsations in nature, the numerical
modelling of this type of pulsation had remained a serious challenge until
recently.  In fact, it had become abundantly clear that purely radiative
models, \ie models that disregarded convective transport, were not capable of
yielding DM pulsations, except on a purely transient basis, \ie they were
switching from one mode to another, but too fast to account for the observed
fairly steady nature of DM pulsators.  Actually, some purely radiative models
of RRd stars had been found already by Kov\'acs and Buchler (1993), but those
results were not satisfactory.  However, recently, DM behavior has been found
simultaneously, and fully independently, in RR~Lyrae models by Feuchtinger
(1998) and in Cepheid models by Koll\'ath \etal (1998), this with different
numerical methods, \viz the Vienna and the Florida codes. The breakthrough came
as a result of the inclusion of time-dependent turbulent convection in the
models.

\section{Amplitude Equations -- Simple Models for DM Pulsations}
\label{sectae}

Numerical hydrodynamics may provide us with an accurate description of the
pulsations of an individual stellar model, but it does not give us an overview
of the behavior when model parameters are changed.  On the sole basis of
numerical hydrodynamical modelling one often has no clue why neighboring models
may exhibit very different pulsational behavior.  Examples of such sensitive
behavior will be discussed below.

Therefore a global way had to be devised for gaining a view of the global modal
selection picture (or the {\sl bifurcation diagram} as it would be called more
generally).  The simplest global way to describe the interaction of pulsation
modes, including DM behavior, is through the use of {\it amplitude equations},
which are also known as normal forms. A derivation of the amplitude equation
formalism applied to stellar pulsation can be found elsewhere (Buchler \&
Goupil 1984, Buchler \& Kov\'acs 1986, \cf also Dziembowski \& Kov\'acs (1984),
Takeuti (1985), Goupil \& Buchler 1994 and for a review Buchler
1993).  This formalism is based on dynamical systems theory, and it is very
general and fundamental.  The amplitude equations give the temporal behavior of
the amplitudes and the phases of the modes that are involved in the nonlinear
pulsation.  They are of fundamental importance for understanding the changes in
the morphology of the Fourier decomposition coefficients of the light and
radial velocity curves as a function of pulsation period.  For example, they
explain the nature of the well-known Hertzsprung progression of the classical
Cepheids.  The amplitude equations show unambiguously how these phenomena are
related to and caused by internal resonances of the pulsating mode with an
overtone (for a review \cf Buchler 1993).

We consider here the situation in which only two modes are involved in the
pulsation and there exists no resonance condition between them nor with the
other modes of the star.  This is in contrast to earlier purely radial
modelling which had required the presence of resonances (Kov\'acs \& Buchler
1993), but which turned out to be unsatisfactory.

We denote the linear eigenvalues by $\sigma_j$, for an assumed
$\exp(\sigma t)$ dependence,

\begin{eqnarray}
\sigma_j &=& \kappa_j+i\omega_j\\ P_j &=& 2\pi / \omega_j\\ \eta_j &=& 2
\kappa_j P_j
\end{eqnarray}

The absence of resonance means that there is no relation of the form $n_1
\omega_0 + n_2 \omega_1 \approx 0$ or $n_1 \omega_0 + n_2 \omega_1 + n_2
\omega_3 \approx 0$, where $n_1$, $n_2$, $n_3$ are small positive or negative
integers.

The generic amplitude equations, appropriate to the nonresonant situation, when
truncated at 5th order read (\eg Buchler 1993):

\begin{eqnarray}
\dot a_0& =   &a_0 (\sigma_0 + Q_{00} \vert a_0\vert^2 
                         + Q_{01} \vert a_1\vert^2
                         + S_{0} \vert a_0\vert^2 \vert a_1\vert^2\nonumber\\
                         &~&~~+ R_{00} \vert a_0\vert^4
                         + R_{01} \vert a_1\vert^4) \\
\label{cmplxaes1}
                                                    \hfill \nonumber    \\
\dot a_1&  =  &a_1 (\sigma_1 + Q_{10} \vert a_0\vert^2 
                         + Q_{11} \vert a_1\vert^2
                         + S_{1} \vert a_0\vert^2 \vert a_1\vert^2\nonumber\\
                         &~&~~+ R_{10} \vert a_0\vert^4
                         + R_{11} \vert a_1\vert^4)
\label{cmplxaes2}
\end{eqnarray}

\vskip 5pt

\ni The $a_j$'s are the complex amplitudes of the two excited modes, and
$Q_{jk}$, $S_j$ and $R_{jk}$ are the complex nonlinear coupling constants.

Note that only selected powers of the amplitudes appear in
Eqs. (\ref{cmplxaes1},\ref{cmplxaes2}); for example there are no quadratic
terms.  Normal form theory (\eg Coullet \& Spiegel 1983) shows (a) that  these
equations are generic, \ie they apply to any system that is characterized by
having two excited nonresonant modes, be it a pulsating star, a biological
system, or any other system, in which the relative growth-rates of the modes
are small ($\eta_j \ll 1$); and (b) that the omitted terms are not essential
for the modal selection (bifurcation diagram), \ie that the {\it nature} of the
dynamical behavior does not depend on them.

Usually it is more convenient to use real amplitudes $A_j(t)$ and phases
$\varphi_j(t)$, defined by $ a_j(t) = A_j(t) \exp i\varphi_j(t)$.  Inserting
this relation into the above equations one gets:
 \begin{equation}
 \dot A_j = A_j\th(\kappa_j + q_{j0} A_0^2 + q_{j1} A_1^2 + s_{j} A_0^2 A_1^2 +
                         r_{j0} A_0^4 + r_{j1} A_1^4)
 \label{aes}
 \end{equation}
 and
 \begin{equation}
 \dot \varphi_j = \omega_j + \hat q_{j0} A_0^2 
                          + \hat q_{j1} A_1^2
                          + \hat s_{j} A_0^2 A_1^2
                          + \hat r_{j0} A_0^4
                          + \hat r_{j1} A_1^4,
\label{phes}
\end{equation}

\ni where we have introduced the real constants by the relations: 
\begin{eqnarray}
Q_{jk}   &=& q_{jk}+i\hat q_{jk}\\
S_{j}    &=& s_{j}+i\hat s_{j}\\
R_{jk}   &=& r_{jk}+i\hat r_{jk}
\end{eqnarray}

\ni One notes that the introduction of the real amplitudes $A_j$ in
Eqs.~(\ref{cmplxaes1}, \ref{cmplxaes2}) produces a complete decoupling of the
amplitudes from the phases.

In principle, the constants $Q$, $S$ and $R$ can be derived from the stellar
structure equations and the linear eigenvectors as was shown in Buchler \&
Goupil (1984).  In practice, however this is a daunting task that has only been
attempted in some very specific cases (Takeuti \& Aikawa 1981), Klapp \etal
1985, Dziembowski \& Krolikowska 1985).  An alternative approach that works
quite well is to derive them from numerical hydrodynamical studies (\eg Buchler
\& Kov\'acs 1987, and below).

In the earliest use of amplitude equations for describing DM pulsations
(Buchler \& Kov\'acs 1986) the amplitude equations were truncated at the lowest
nontrivial, \ie cubic, terms.  It was shown that in this order the single-mode
fixed points and steady DM pulsations cannot simultaneously be stable for the
same stellar model.  Subsequently, the behavior of the hydrodynamical models and
the observational constraints imposed by the Beat Cepheids and RR Lyrae stars
have forced us to consider also the next order, quintic terms (Buchler, Yecko,
Koll\'ath \& Goupil 1999, hereafter BYKG).  The properties of these equations
and their applicability to the DM stellar pulsators were discussed in that
paper.  Here we present a slightly different discussion and approximation.
While, for simplicity, BYKG retained only the $r_{00}$ and $r_{11}$ cubic
terms, instead, we keep the $s_0$ and $s_1$ cubic terms because they are
sufficient to give a good, and in fact a better fit to the results of the
hydrodynamical calculations.

\subsection*{Time-Independent Amplitude Equations}

If our primary interest is the study of steady pulsations, \ie oscillations
with constant amplitudes, then we need to consider only the fixed points (FP's)
of the amplitude equations Eq.~(\ref{aes}), defined by $\dot A_0 = \dot A_1 =
0$.  When only one FP amplitude is nonzero (a single-mode FP) the corresponding
full-amplitude (nonlinear) pulsations of the star have constant amplitudes and
they are mono-periodic.  These pulsations are then called limit-cycles, when in
addition they are stable.

Eqs.~(\ref{phes}) show that for FP's the phases become linearly increasing
functions of time (because the $A_j$'s are constant) and they thus yield the
nonlinear modal frequencies $\bar\omega_j = \dot\varphi_j$.

The possible FP's are therefore obtained by looking for the solutions
of the 4 combinations of the equation pairs:

\begin{eqnarray}
\kappa_0 + q_{00} A_0^2 + q_{01} A_1^2 + s_{0} A_0^2 A_1^2 +
                         r_{00} A_0^4 + r_{01} A_1^4& = &0  
\label{fpae0}
\\
A_1& = &0 
\label{fpae1}
\end{eqnarray}
\begin{eqnarray}
\kappa_1 + q_{10} A_0^2 + q_{11} A_1^2 + s_{1} A_0^2 A_1^2 +
                         r_{10} A_0^4 + r_{11} A_1^4& = &0 
\label{fpae2}
\\
A_0& = &0 
\label{fpae3}
\end{eqnarray}
Without loss of generality, we label the two modes 0 and 1, because most of the
time we will be talking about the fundamental and first overtone modes.

Eqs.~(\ref{fpae0} -- \ref{fpae3}) are a function of the squared amplitudes. It
is therefore sometimes advantageous to work in ($A_0^2,A_1^2$) space.  Thus in
Figure~\ref{figeoe1bif} we have plotted the loci defined by the first of each
of Eqs.~(\ref{fpae0}, \ref{fpae2}) as solid and dotted lines,
respectively, \th for
an RR Lyrae model that will be discussed below.  The remaining two loci are
the positive x and y axes.  The solid and open circles represent stable and
unstable FPs which are the intersections of these loci.  Note that the
situation is much simpler when all the quintic terms are disregarded as in
Buchler \& Kov\'acs (1986) because then the loci degenerate into a set of
straight lines, and clearly there can be at most one DM solution in that case.
Here we consider the more general case.

\begin{table}[ht]
 \caption{Possible fixed point structures of the nonresonant amplitude
equations (in addition
to the unstable trivial FP, $A_0=A_1=0$), 
assuming $\kappa_0$, $\kappa_1>0$.  S: stable, U: unstable.}
\begin{tabular*}{\hsize}{@{\extracolsep{\fill}}cccccc}
\hline 
 Number of FP's & F & O$_1$ & DM1 & DM2\cr
\hline 
   2  & S  & U  & -- & -- \cr
   2  & U  & S  & -- & -- \cr
   3  & S  & S  & U  & -- \cr
   3  & U  & U  & S  & -- \cr
   4  & S  & U  & S  & U  \cr
   4  & U  & S  & S  & U  \cr
\hline
\end{tabular*}
\label{tabfp}
\end{table}

The trivial FP of the amplitude equations is the static stellar model, \ie with
both $A_0=0$ and $A_1=0$. If both growth-rates are negative ($\kappa_0,
\kappa_1<0$) the FP is stable, and so is the star.  The next simplest case is
where only one $\kappa$ is positive in which case the only possible pulsation
state of the star is a limit-cycle in that mode.  For the existence of a DM
solution both growth-rates need to be positive, but as we will see it is not a
sufficient criterion for steady DM oscillations.  With the nonresonant
amplitude equations only a limited number of possible limit-cycle combinations
are possible, and they are listed in Table~\ref{tabfp}.

\begin{figure}[ht]
\centerline{\psfig{figure=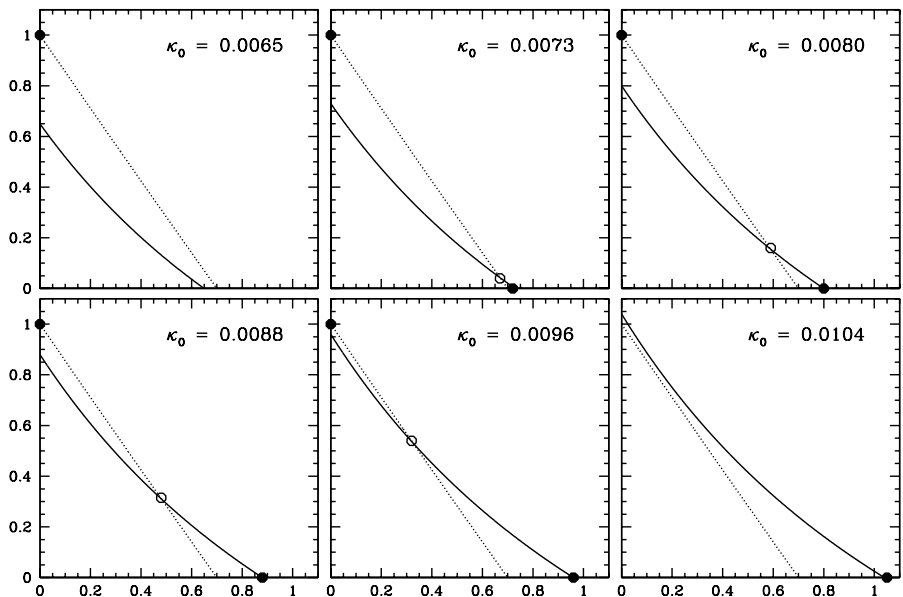,width=12cm}}
\caption{DM solutions of the amplitude equations in the ($A_0^2,A_1^2$ 
plane for an RR Lyrae model.  Solid circles are stable and open circles are
unstable fixed-points}
\label{figeoe1bif}
\end{figure}

\begin{figure}[ht]
\centerline{\psfig{figure=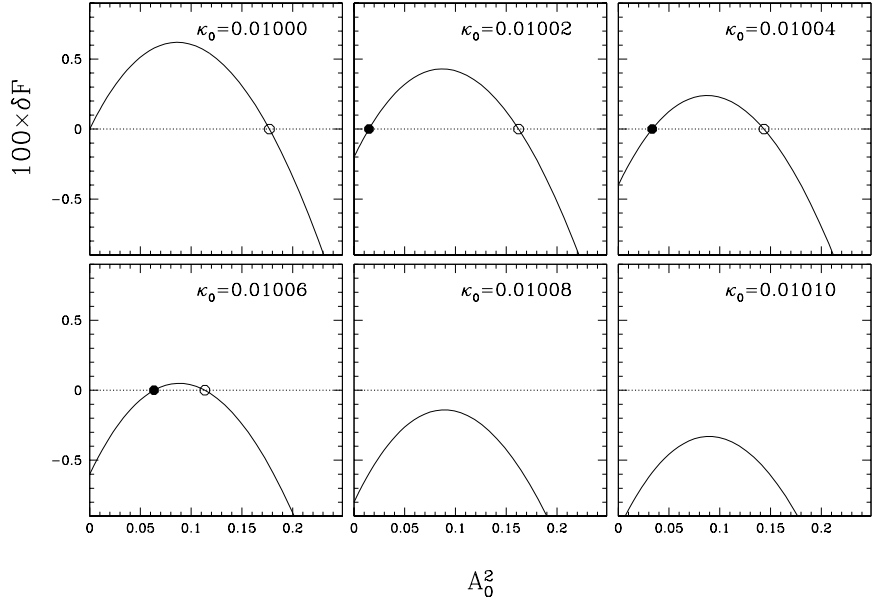,width=11.5cm}}
\caption{DM solutions of the amplitude equations, see text Eq.~(\ref{Feq});
Solid circles denote stable and open circles denote unstable DM
solutions.}
\label{figsq}
\end{figure}

In order to make the problem tractable we now keep the $s$ terms, but omit the
$r$ quintic terms in Eq.~(\ref{fpae0},\ref{fpae2}).  
It is simple to find the
fundamental and first overtone limit-cycle amplitudes

\begin{eqnarray}
A^2_{0,lc} &=& {-\kappa_0\over q_{00}},  \quad \quad A_{1,lc} = 0 \\
A^2_{1,lc} &=& {-\kappa_1\over q_{11}},  \quad \quad A_{0,lc} = 0 .
\end{eqnarray}

\ni Inserting those values into the amplitude equations corresponding to the
other mode, we can easily obtain the conditions for the linear stability of the
fundamental and overtone limit-cycles, respectively:
\begin{eqnarray}
\bar\kappa_1 \equiv \kappa_1 + q_{10} A_{0,lc}^2  
            = \kappa_1 - \kappa_0 {q_{10} \over q_{00}} < 0 \\
\bar\kappa_0 \equiv \kappa_0 + q_{01} A_{1,lc}^2  
            = \kappa_0 - \kappa_1 {q_{01} \over q_{11}} < 0
\end{eqnarray}

We illustrate the modal behavior with an RR Lyrae model for which we have
computed the nonlinear coupling coefficients (see Table~\ref{tabae}).  The
numbers have been rounded and the $q$'s and $s$'s have been normalized to set
the limit-cycle amplitudes to the unity (\eg $q_{00}=-\kappa_0$ and
$q_{11}=-\kappa_1$).

In order to see the behavior of the solutions along a sequence of RR Lyrae
models, we fix all these parameters but let $\kappa_0$ vary.  This sequence of
solutions, obtained from Eqs.~(\ref{fpae0} -- \ref{fpae3}), has already been
depicted in Fig.~\ref{figeoe1bif}.

From the stability of the limit-cycles one further finds that the range of
stable fundamental mode pulsations is $\kappa_0>0.007$.  The range for the
stable overtone is $\kappa_0<0.01$. This indicates that both limit-cycles are
possible (the 'either-or' regime in the pulsation jargon) and these multiple
solutions exist in the $0.007<\kappa_0<0.01$ interval, and necessarily there
is a DM fixed point for that regime which is furthermore unstable.

In Figure~\ref{figeoe1bif} one easily misses some important and interesting
solutions that occur in an extremely narrow range of $\kappa_0=0.0100$ to
0.0101, located between the last two subfigures.  In fact for the parameters of
Table~\ref{tabcc} the upper, solid curve undergoes two intersections with the
dotted curve which corresponds to two additional DMs.

In order to exhibit this behavior in a more apparent way we consider the
quantities $F_j$, derived from Eqs.~(\ref{fpae0}, \ref{fpae2}).

\begin{equation}
F_j = A_1^2 = - {\kappa_j + q_{j0} A_0^2 \over  q_{j1}  + s_j A_0^2} 
  \quad\quad j=0,1 .
\label{Feq}
\end{equation}

Plots of $\delta F \equiv F_1-F_0$ as a function of $A_0^2$ give the amplitudes
of the DMs solution.  It can be seen than the fixed point is stable if the
slope of $F_0$ is steeper than that of $F_1$, \ie $d\th \delta F / d A_0^2 >
0$.  Otherwise no stationary DM pulsation is possible.

A stable DM solution exists and both limit-cycles lose their stability if and
only if $\delta F$ is an increasing function of $A^2_0$ in the neighborhood of
the fixed point.  With the opposite slope both the fundamental and the first
overtone limit-cycles are stable, giving the ``either-or'' region of pulsation
(\cite{BK86}).

Figure~\ref{figsq} shows how the $\delta F = F_1-F_0$ function varies as
$\kappa_0$ is changed. Only a very narrow range exists where the nearly
parabolic function has two intersections with the zero line.

\begin{table}[ht]
\caption{Growth-rates and nonlinear coupling coefficients from an RR Lyrae 
model.}
\begin{tabular*}{\hsize}{@{\extracolsep{\fill}}cccccccc}
\hline 
$\kappa_0$ & $\kappa_1$ & $q_{00}$ & $q_{10}$ & $q_{01}$ & $q_{11}$ 
& $s_0$ & $s_1$ \cr
\hline 
0.01 & 0.042 & --0.01 & --0.06 & --0.01 & --0.042 & --0.006 & --0.001 \cr
\hline
\end{tabular*}
\label{tabae}
\end{table}

Both Figs.~\ref{figeoe1bif} and \ref{figsq} show that the DM amplitudes are
always smaller than the corresponding SM limit-cycle amplitudes.  When the DM
occurs near the $A_1$ axis, then $A_{1,dm}\leq A_{1,lc}$, but $A_{0,dm} \ll
A_{0,lc}$.  With the given parameters the maximum fundamental amplitude is only
30\% of the limit-cycle amplitude. For most of the RRd stars the fundamental
mode amplitude is indeed smaller than that of the first overtone (see Kov\'acs
in this Volume), indicating that the above RR~Lyrae model is on the right
track.

\subsection*{Time-Dependent Amplitude Equations}

The time-dependent solutions of the amplitude equations become essential when
we interpret the numerical hydrodynamical integrations or explain the changes
due to \eg evolutionary changes of the star. As we see above, very small
changes in $\kappa$'s can induce significant changes in the possible pulsation
state of the star. In our test problem if the fundamental mode growth-rate is
slowly increasing, at $\kappa_0=0.01$ the overtone mode loses its stability.
Then the model evolves smoothly to the DM state, if the system has been
pulsating in the overtone mode before.  With a further increase of $\kappa_0$,
at $\kappa_0\approx0.01007$, the fundamental mode pulsation remains the only
stable state.  Then the amplitude equations can be used to estimate how the
stars switches from the DM state to the fundamental limit-cycle.

In Fig.~\ref{figms} the temporal variation of the amplitudes are shown during
the simulated mode switching.  The integration of the amplitude equations was
initiated (at t=0) at the stable DM fixed point, and $\kappa_0$ was
continuously increased at a $10^{-7}$ years${}^{-1}$ rate.  The DM oscillations
became unstable during the first 10 years, but it took more than 150 years to
evolve the star to the fundamental mode pulsations.
 
Right at the bifurcation point where the stable DM pulsation trades stability
with the unstable fundamental SM limit-cycle the growth-rate of the DM
pulsation is infinitely small and it takes evolution for the star to move into
the regime where this growth-rate achieves sufficiently large values for the
switch to the fundamental limit-cycle to occur.  It is therefore clear that the
mode switching time-scale depends both on the evolution time-scale and on the
sensitivity of the modal growth-rate to the evolution time-scale.

\begin{figure}[ht]
\vskip 10pt
\centerline{\psfig{figure=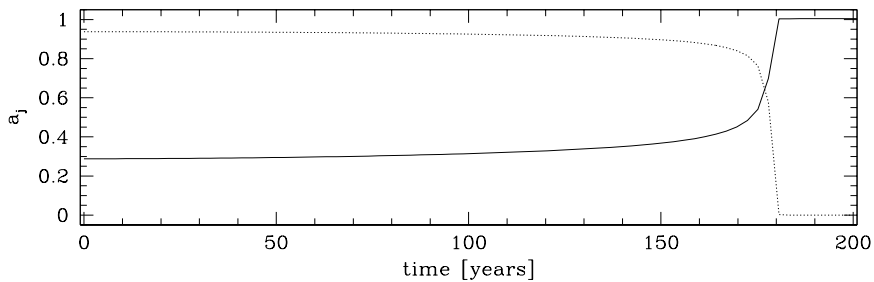,width=11.5cm}}
\caption{Simulated mode switching as predicted by the amplitude equations, see
text; solid: $A_0$, dotted: $A_1$.}
\vskip 10pt
\label{figms}
\end{figure}

In section \ref{sectres} we discuss how the hydrodynamical calculations can be
used to reconstruct the amplitude equations.

\section{Hydrodynamical Modelling of Double-Mode pulsation}

\subsection*{The Hydrodynamical Code}

The classical variable stars that we consider in this review are radial
pulsators.  The global motions thus can be modelled by 1D hydrodynamics (see
however Kov\'acs in this Volume for indication of nonradial modes in RR Lyrae
stars).  Since real 3D modelling remains a dream in nonlinear pulsation
calculations, the local nonradial flows \ie turbulence have to be treated in
some one dimensional approximation.  Recipes and pulsation codes for 1D
turbulent convection (TC) have been developed by different authors (\eg
Stellingwerf 1982, Kuhfu\ss~1986, Gehmeyr \& Winkler 1992, Feuchtinger,
M. U. 1998, Yecko, Koll\'ath, \& Buchler 1998).

The fluid dynamics part of the model calculations are given by the
following equations:

 \begin{eqnarray}
  {du\over dt} &=& -{1\over\rho}{\partial \over\partial r}
\left(p+\th\th p_t+p_\nu\th\th \right)
   - {G M_r\over r^2} \quad\quad \\
 & & \nonumber \\
   {d\th e\over dt} +p\th  {d\th v\over dt}
  &=& -{1\over\rho r^2} {\partial \over\partial r} \left[ r^2
\left(F_r+\th\th F_c\th\th \right)\right]  +\th \mathbf{{\cal C}}
 \quad\quad
 \end{eqnarray}                                                            

The turbulent motion of the gas and the convection interacts with the
hydrodynamics of the radial motion through the convective flux $F_c$, the
viscous eddy pressure $p_\nu$ the turbulent pressure $p_t$, and finally,
through an energy coupling term ${\cal C}$.  Those terms can be derived by
averaging the 3D hydrodynamics equations.  This procedure introduces a sequence
of moment equations (see \eg \cite{canuto,canutod}).  In the simplest recipe we
truncate all this equations, but leave the time-dependent diffusion equation
for the turbulent energy $e_t$:

 \begin{equation}
   {de_t\over dt} +
    \left(p_t+p_\nu\right)\th  {d\th v\over dt}
  = -{1\over\rho r^2} {\partial \over\partial r}\left( r^2 F_t\right)
   -{\cal C} \quad\quad \nonumber\\
 \label{eqs_tc}
\end{equation}

The coupling term that connects the gas and the turbulent energy
equations is given by:
 \begin{equation}
 {\cal C}=   - { e^\ah_t\over\Lambda}
  \alpha_d\th \left(e_t - S_t\right),
 \end{equation}
where $\Lambda = \alpha_\Lambda\th H_p$, $H_p = p\th
r^{\scriptstyle 2}/(\rho GM)$ is the pressure scale height
$\alpha_\Lambda$ is the mixing length parameter 
and $\alpha_d$ is a dimensionless parameter.

Both the convective flux and the source term of the turbulent energy ($S_t$)
depend on the entropy gradient:

 \begin{equation}
 Y = -{H_p\over c_p} \th\th {\partial s\over \partial r}
 \end{equation}

\begin{equation}
  S_t = (\alpha_s\th \alpha_\Lambda)^{\scriptstyle 2} \th\th
      {p\over \rho} \th\beta T\th\th Y \, f_{pec},
 \label{eqgw}
\end{equation}

\begin{equation}
  F_c = \alpha_c\alpha_\Lambda\th \rho e_t^{\ah} \th  
        c_p T\th  Y \, f_{pec},
\end{equation}

\ni where $\beta$ is the thermal expansion coefficient $\alpha_s$, $\alpha_c$
are parameters and other symbols have their usual meanings.  The P\'eclet
correction $f_{pec}$ accounts for the decrease of convective efficiency when
radiative losses are important.  We approximate this correction factor by

\begin{eqnarray}
 f_{pec}   &=& {1\over 1+\alpha_r P\ngth e^{-1}} \\
 P \ngth e &=&  D_c/D_r \\
       D_r &=& {4\over3} {a c T^3 \over \kappa\rho^2 c_p} \\
       D_c &=& \Lambda e_t^\ah 
\end{eqnarray}

The remaining quantities are defined as

\begin{eqnarray}
  p_t   &=& \alpha_p \th \rho \th e_t   \\
  p_\nu &=& - \frac{4} {3} \alpha_\nu\th \rho \Lambda   e^\ah_t  
                  r{\partial\th \over\partial r}{u\over r} \\
  F_t   &=& -\alpha_t\th \rho \Lambda  e^\ah_t {\partial\th\th e_t
  \over\partial r}
\end{eqnarray}

The TC model equations are based on a very simple physical picture of turbulent
energy generation and diffusion, and convective energy transport.  The various
expressions can essentially be derived from dimensional arguments (even though
numerous attempts have been made to derive them from more detailed analyses -
for a dated, but still excellent review, see Baker 1987).  The above
definitions of the turbulent energy source and of the convective flux are not
unique, and slightly different expressions have been used by different
authors. The formulation adopted is of that of Gehmeyr \& Winkler (1992).  For
a comparison of the different recipes used for stellar pulsation calculations
see \cite{BYKG}.

The dimensionless $\alpha$ parameters of our turbulent convection model
equations are of order unity, but theory provides no guidance as to their
numerical values.  We will resort to a comparison of numerical results with
empirical data, such as the position and the width of the instability strip, to
calibrate those parameters.

We have not yet performed a full calibration, which is a giant task because of
the number of parameters, and in fact it is not even clear whether all
observational constraints can be fulfilled with the use of such a simple model
for turbulent convection.  In the meantime we have chosen several sets of
$\alpha$ parameters that satisfy some of the constraints, and we have explored
the consequences of turbulent convection for the linear and nonlinear
properties of pulsation.

It has been comforting that, for the first time, we have been able to calculate
DM pulsation with these parameters for a wide set of (physical) stellar
parameters. However the results still miss some of the observational
constraints, \eg the model amplitudes are small compared to the observed ones,
the modelling of 1st overtone/2nd overtone DM pulsations is still not quite
satisfactory.  At this time no detailed or comprehensive comparison has been be
made between the observations and the DM modelling.  Thus this paper is a
status report on modelling rather than a closed and self contained theory of DM
pulsations.  However, the tendencies, \eg along a sequence with different
temperatures, can be used as primary tests for our models.

The first step in the numerical modelling is the construction of the static
envelope model, and its linear stability analysis which yields the possible
pulsation periods and the corresponding excitation-rates.  These modal
growth-rates tell us only the rate of increase or decrease of the oscillation
for tiny amplitudes.  Furthermore, if more than one mode is excited we cannot
infer from this information which of the modes will grow into full-amplitude
nonlinear oscillations.  This 'modal selection' problem has already been
discussed in section~\ref{sectae} where it was shown how it can be attacked
with the amplitude equation formalism.

There exists no really simple way for computing the limit-cycles or the final
DM or multi-mode pulsation state of the stellar models. The hydrodynamical
modelling is an initial value problem, and we need to evolve the model directly
with the hydrodynamics code, or through a relaxation method to obtain the
stable oscillations.  Furthermore, the model can evolve to different final
stationary states depending on the initial kick of the static envelope.
Fortunately, the investigated models have only one or two possible stable
pulsation states, since for radially pulsating envelopes only a few modes are
generally linearly unstable (at least for the Cepheids and RR~Lyrae).  In the
case of nonradial pulsations, where dozens of modes are linearly unstable, the
number of stable NL oscillations can be higher.

A great advantage of the Florida code is that it can relax to the periodic
solutions by iteration and it provides the stability (Floquet) analysis of the
limit-cycles. When both the fundamental mode and the first overtone are
linearly unstable but the other modes are stable, then the Floquet exponents of
their limit-cycles provide the primary information on modal selection.

\subsection*{DM Solutions in the Numerical Calculations}

The computation of the possible full-amplitude (or nonlinear) periodic
pulsation states of a stellar model is essentially routine work with the
relaxation code.  However there exists no similar method for finding DM or
other multi-mode full amplitude pulsations.  Even when the fundamental and the
first overtones are both linearly unstable, nothing guarantees the existence of
steady DM pulsations.  If both periodic limit-cycles are unstable against
perturbations, then the model should oscillate in a DM state only.  However, as
the solutions of the amplitude equations indicate (see Section~\ref{sectae})
stable DM solutions can exist even when one of the single-mode limit-cycles is
stable.  In such a case the Floquet analysis is insufficient for identifying DM
pulsations.

The most efficient way of kicking the initial static model with the velocity
eigenvectors from the linear stability analysis.  In order to obtain different
transient paths we thus use linear combinations of the fundamental and the
overtone mode eigenvectors with a surface amplitude of $\approx10$ km/s.

\begin{figure}[ht]
\centerline{\psfig{figure=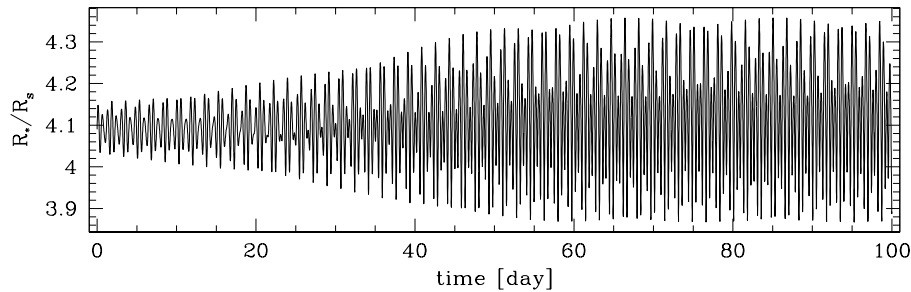,width=12cm}}
\caption{ The initial transient in RR Lyrae model pulsations}
\label{figrrh}
\end{figure}

In Figure~\ref{figrrh} the radius variation of an RR Lyrae model is shown for
the initial 100 days of model integration. The multi-component nature of the
signal can be seen from an inspection of the curve -- the beats are clearly
visible. The amplitudes of the modes are seemingly saturated, but further
integration of the model shows that this is not the case.  At that stage of the
hydrodynamical evolution one cannot predict the final pulsation state: it can
be a limit-cycle with any of the unstable modes, or a DM pulsation.  The system
generally has to be integrated for a {\sl very long} time (usually for
thousands of cycles) to determine whether the amplitude of the modal components
saturates at some finite value.

To emphasize the importance of long term integration we have calculated the
amplitudes from an extended continuation of the transient shown in
Fig.~\ref{figrrh}.  The variation of the amplitudes is displayed in
Fig.~\ref{figtr}. Note the different time-scales on the two figures: the curve
shown in Fig.~\ref{figrrh} corresponds to the very beginning of the amplitude
evolution on Fig.~\ref{figtr}.

\begin{figure}[ht]
\centerline{\psfig{figure=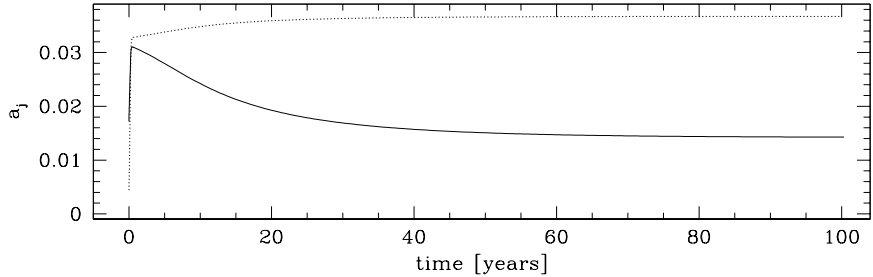,width=11.5cm}}
\caption{ The evolution of the amplitudes after the transient shown
in Fig.~\ref{figrrh}. Solid line: fundamental mode, dotted line: first overtone.}
\label{figtr}
\end{figure}

Claiming DM behavior is therefore not an easy task. Very long transients with
mixed-mode behavior exist in the hydrodynamical calculations, even when the
model finally converges to a SM limit-cycle. This means that the knowledge of
the Fourier spectra and of the velocity and light-curves from a single
hydrodynamical calculation of a model cannot provide a sufficient proof of
steady DM behavior.  A more elaborate set of calculations is necessary.  This
is the topic of the next sections.

\subsection*{Time-Dependent Amplitudes and Phases}

Frequently a visual inspection of the time-series that is obtained from the
hydrodynamical calculations may indicate that the pulsations have reached a
steady state.  However these indications can be quite treacherous.  Even the
Fourier analysis of the light, velocity or radius variation can be misleading
because of the temporal changes of amplitudes and periods in the transitory
stage of model calculations.  In order to obtain reliable information
from the numerical results it is necessary to apply a well suited
time-frequency analysis and derive instantaneous amplitudes and periods.

A variety of such methods are available.  For example, Kov\'acs, Buchler
\& Davis (1987) used a time-dependent Fourier method.  They performed linear
least-squares fits with sine functions to very small successive portions of the
data. The bases of the successive fits were shifted by small portions of the
fit, and the results were averaged to get smooth results. The disadvantage of
this method is that short bases provide large errors on the amplitudes, while
longer bases introduce temporal averaging destroying the instantaneous nature
of the amplitudes and phases.

Instead of the time-dependent Fourier analysis one can follow G\'abor (1946) to
reconstruct the time-dependent amplitudes and frequencies of the pulsation
modes.  Let $s(t)$ represent the real part of an assumed complex analytical
function $a(t)$.  The imaginary part $\tilde s(t)$ of $a(t)$ can then be
obtained via a Cauchy integral, which through contour deformation becomes a
Hilbert transform.  Physicists are generally familiar with this analytic signal
concept through the Kramers--Kronig dispersion relations (Jackson 1975).  

\begin{eqnarray}
a(t) 
&=& s(t)+i \tilde s(t) = s(t)+\frac{i}{\pi } 
p.v. \int_{-\infty}^\infty \frac{s(t')}{t-t'}dt' \\
&=& \frac{1}{\pi } \int^{\infty }_{0}
          \int_{-\infty}^\infty s(t')e^{i\omega (t-t')}dt'd\omega \\ 
&\equiv& A(t)e^{i\varphi (t)} 
\end{eqnarray}
It is easy to verify for example that with $s(t)= A \cos \omega t$ one finds
$\tilde s(t) = A \sin \omega t$.  The G\'abor construction makes it possible to
unambiguously define the phase $\varphi(t)$ and the amplitude $A(t)$ of a
signal.  Consequently it allows one to extract the instantaneous frequency
(Cohen 1994).

It is possible to extend the method to multi-component signals.  To obtain the
instantaneous amplitudes and periods of the pulsation modes separately one
first filters the signal to eliminate the power from the other frequency
components. It is convenient to make the filtering in Fourier space, and
combine it with the definition of the analytic signal

 \begin{equation} 
 Z_{k}(t) = a_{k}(t)e^{i\varphi_{k}(t)}
          = \frac{1}{\pi }\int ^{\infty }_{0} H(\omega -\omega _{k})
          \int_{-\infty}^\infty s(t')e^{i\omega (t-t')}dt'd\omega,
 \label{eqaf} 
\end{equation}
where $H(\omega -\omega _{k}) $ is the window of the filtering, centered on
$\omega_k$.  A Gaussian window with a half width of 0.2 c/d gives a
satisfactory result for RR Lyrae models. The resulting amplitude and phase
gives the temporal evolution of the given mode. The least-squares fitting of a
Fourier sum to the radius variation gives essentially the same results. However
the analytic signal gives better resolution and it is not necessary to average
the amplitudes in time to get smooth results. Thus it is more appropriate for
further analysis. A great advantage of the analytic signal is that the
resulting amplitudes can be directly used to fit the coefficients of the
amplitude equations in a differential form -- a simple linear least-squares
method. The amplitudes calculated by the method of KBD87 are not suitable for a
differential fit, but only for a complicated nonlinear least-squares fit to the
integral curves.

\begin{figure}[ht]
\centerline{\psfig{figure=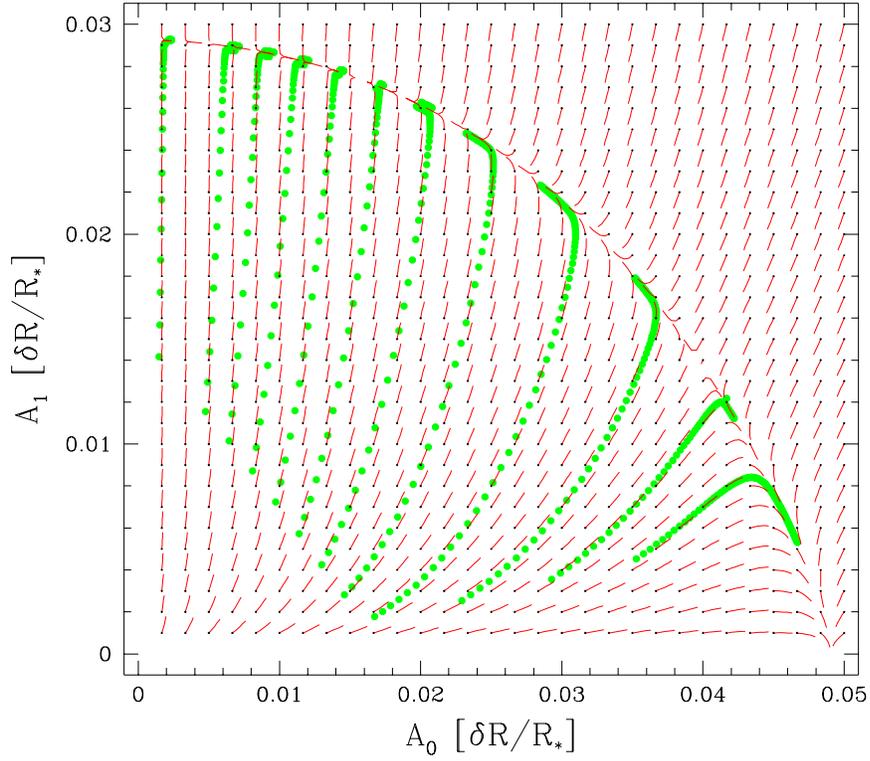,width=11.5cm}}
\caption{ The flow in the ($A_0,A_1$) phase-space. 
The large dots represent the hydrodynamical results. 
The short lines represents the flow in the phase-space.  The two separatrices
are clearly visible: the first runs from the origin upward between the
fourth and fifth evolutionary tracks, and the second one is the arc 
connecting the two single-mode fixed points.}
\label{figflow}
\end{figure}

\subsection*{Modal Selection Problem}

Armed with this tool we can now analyze the modal selection problem in the
hydrodynamical models.  A time consuming, but very reliable method is to
perform a number of hydrodynamical integrations with different initial
perturbations of the same static model, and then to extract the slowly varying
amplitudes with the help of a time-frequency analysis.  The resulting
phase-portrait ($A_1(t)$ vs. $A_0(t)$) gives a clear global overview of the
nonlinear behavior of the model, and thus permits one to see immediately if
stable DMs exist.

In Fig.~\ref{figflow} we show a representative set of such hydrodynamical
tracks for a Cepheid model.  The initial perturbations have been consisted of a
mixture of the fundamental and the first overtone eigenvectors.
Fig.~\ref{figflow} provides conclusive proof of the presence of stable beat
pulsation with amplitudes $A_0 = 0.016$ and $A_1 = 0.028$ that coexists with
the stable fundamental mode with amplitude $A_0 = 0.049$.  In this case there
is hysteresis and the actual state of the star depends on its former
evolutionary path.  From Fig.~\ref{figflow} one can also infer that two
additional nontrivial fixed points exist for the given model in the phase-space
of the amplitudes, namely a first overtone limit-cycle and a DM saddle-point,
but these states must be unstable.  The observed behavior of the model is in a
good agreement with the results found with the amplitude equations.  The
amplitude equations can be used to find all the fixed points, even if the
hydrodynamical trackss do not actually reach them, either because of
insufficient integration time, or because they are unstable and thus simply
unreachable this way.  The flow field that is obtained from the amplitude
equations is represented by the short thin lines in Fig.~\ref{figflow}.

The procedure that we have described, namely numerical hydrodynamics
integrations with judiciously chosen different initial conditions, combined
with a time-frequency analysis and with a Floquet analysis provide a complete
picture of the possible pulsation states of a given model and of their
stability.  By repeating the calculations with different effective temperatures
and luminosities one can thus map the whole picture of modal selection on the
HR diagram.

\section{Results}
\label{sectres}

We have already mentioned that the calibration of the turbulent parameters is a
big task that has not been finished yet.  The results that we present here can
show the general tendencies that exist in the models, but we cannot pretend
that the models can be used to infer for example physical parameters from
individual variable star observations.

\subsection*{The Coupling Coefficients}

The time-dependent amplitude equations describe the evolution of the amplitudes
of the excited modes.  From the knowledge of the transient hydrodynamical
tracks for a given model it is thus possible to evaluate the \apriori unknown
nonlinear coefficients of the amplitude equations.

Of course, in order to extract reliable nonlinear coupling coefficients it is
necessary to integrate the models from a sufficient number of initial
conditions so that the 'trajectories' sample well the phase-phase ($A_0,A_1$).
Numerical simulations have show that multiple DM solutions, a stable fixed
point and a saddle point can coexist in the model pulsations.  From the
analysis of the fixed-points of the amplitude equations one can conclude that
it is necessary to include at least one quintic term into (see
Section~\ref{sectae}).  According to our experiments the mixed quintic
coefficients ($s_j$) are usually sufficient to give a good fit.  The inclusion
of the other quintic terms ($r_{ij}$) results only in a minor improvement of
the fit. We have to note however that fits with the $r_{ii}$ terms only are
also capable of reproducing the main features of the amplitude evolution.

In Table~\ref{tabcc} we present the coefficients of the amplitude equations for
a sequence of RR Lyrae models where a DM solution exists in the $6500K <
T_{eff} < 6515K$ temperature range.  There are no significant variations with
\Teff\ for most of the coefficients, but the growth-rate $\kappa_0$ has a
decreasing trend.

\begin{table}[ht]
\caption{Coupling coefficients in a sequence}
\begin{tabular*}{\hsize}{@{\extracolsep{\fill}}ccccccccc}
\hline 
 $T_{ef\!f}$ & $\kappa_0$&$\kappa_1$&$q_{00}$&$q_{01}$&$q_{10}$&$q_{11}$ 
& $s_0$ & $s_1$ \cr 
\hline 
 6495 & \ngthh 0.009397 & \ngthh 0.03975 &\ngthh  --3.08 &\ngthh  --20.18 
&\ngthh  --10.16 &\ngthh 
     --44.20 & --2789. & --729. \cr
 6500 &  \ngthh 0.009319 & \ngthh 0.03976 &\ngthh  --3.07 &\ngthh  --20.29 
&\ngthh  --10.13 &\ngthh 
     --44.12 & --2788. &  --583. \cr
 6505 &  \ngthh 0.009235 & \ngthh 0.03967 &\ngthh  --3.07 &\ngthh  --20.35 
&\ngthh  --10.13 &\ngthh 
     --43.94 & --2750. & --500. \cr
 6515 &  \ngthh 0.009095 & \ngthh 0.03976 &\ngthh  --3.08 &\ngthh  --20.62 
&\ngthh  --10.08 &\ngthh 
     --43.90 & --2833. & --384. \cr
 6525 &  \ngthh 0.008896 &\ngth  0.03971 &\ngthh  --3.07 &\ngthh  --20.75 
&\ngthh   --9.99 &\ngthh 
     --43.72 & --2792. & -360. \cr
\hline
\end{tabular*}
\label{tabcc}
\end{table}

We have fitted the coefficients with a quadratic function of temperature (\eg
$s_j = C_0 + C_1 T_{eff} + C_2 T_{eff}^2$), and we have used the fit to map the
solution of the amplitude equations as a function of temperature.  Fig~\ref{figseq} shows the
variation of the function $\delta F$ as the temperature is decreased. The
displayed behavior is very similar to the one displayed in Fig~\ref{figsq},
where only one of the parameters, the fundamental mode growth-rate was
changed. If we fix all the parameters but decrease $\kappa_0$ according to the
temperature fit, the basic tendencies on the figures remain the same \ie for
the given example the variation of the fundamental growth-rates plays the most
important role.

\begin{figure}[ht]
\centerline{\psfig{figure=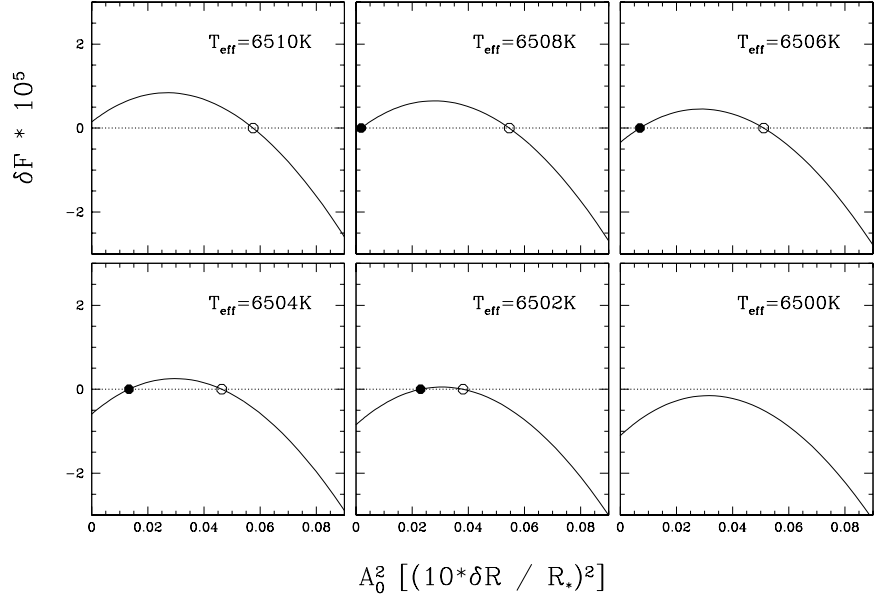,width=11.5cm}}
\caption{Double-mode solutions of the amplitude equations 
fitted to a model sequence.}
\label{figseq}
\end{figure}

\subsection*{The Fourier representation of steady DM pulsations}

The primary information one can derive from observational data is the spectral
content of the light variations, the amplitudes and periods presented in the
Fourier spectra.  (Examples of RR Lyrae frequency spectra can be found in
Kov\'acs's review in this Volume.)  In addition, because of the nonlinearity of
pulsation, linear combinations of the primary frequencies also appears in the
spectra.  A good example for the existence of those terms is the analysis of
the Cepheid TU~Cassiopeia (Szabados 1993), where 29 frequencies have been
identified.  Even terms like 4f$_0$+3f$_1$ or 5f$_0$+2f$_1$ are inferred from
the Fourier transform of the light-curve. In the RR Lyrae spectra typically
only the f$_0$+f$_1$ and f$_0$--f$_1$ terms can be found in addition to f$_0$
and f$_1$.

In order to relate the observed frequency spectra with those calculated from
model pulsations, we have calculated synthetic data with the observed
periodicities but with the same sampling and length as the hydrodynamical
outputs (typically 1000 cycle).  In Figure~\ref{figsp} we compare the spectra
of TU~Cas and of a MACHO DM star presented in Figure 2 of Kov\'acs in this
Volume with typical hydrodynamical results.  We note that the amplitudes are
given in bolometric magnitudes for the models and in 'V' (TU~Cas) and in
instrumental red (MACHO RR Lyr).  The agreement is satisfactory between the RR
Lyrae models and observations.  TU~Cas has a large amplitude variation -- it is
the high nonlinearity that causes the large power at the harmonics of the
fundamental mode.  Our model has a much lower amplitude and consequently a
weaker harmonic power, and it thus misses this behavior.

\begin{figure}[ht]
\centerline{\psfig{figure=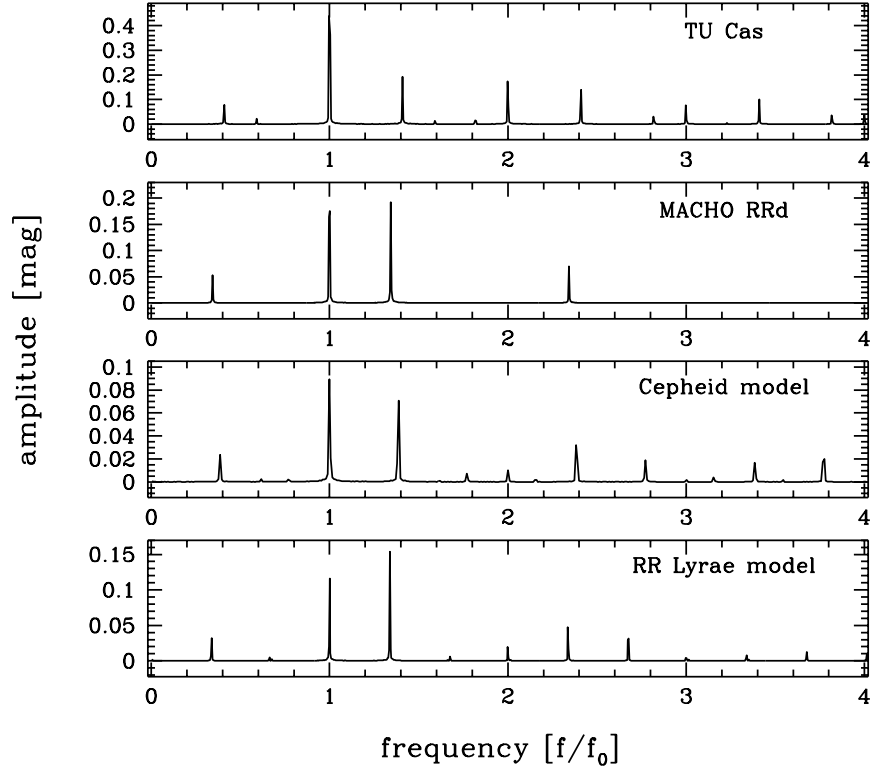,width=11.5cm}}
\caption{Fourier spectrum of the light variation:
From top to bottom: TU~Cassiopeia, MACHO RRd variable,
Cepheid model, RR Lyrae model.}
\label{figsp}
\end{figure}

When the harmonics of the modal frequencies can be observed, the corresponding
Fourier parameters provide useful information.  It has been observed that the
Fourier parameters, $\varphi_{21}$ and $\varphi_{31}$, of both modes in DM
Cepheids are very similar to those of the corresponding modes in mono-periodic
stars with similar periods (\eg Udalski \etal 1999, Beaulieu, private
communication).  Since the models relax very slowly on the separatrix
connecting the F mode to the O$_1$ through the DM solution, one can calculate the
Fourier parameters along the separatrix.  This test provides a clue how the
Fourier phases depend on the modal amplitudes (Fig.~\ref{figfp}).  We have
selected a model from the middle of the DM region for the plot, but we checked
the Fourier parameters for the whole range and obtained very similar
results.  The Fourier parameters were calculated from the hydrodynamical model
light-curves, and the amplitudes are in magnitudes. To indicate the position of
the separatrix, we also show the $A_1$ \vs $A_0$ plot of the models in the top
right box.  As expected the $R_{21}$ amplitude ratios have a strong dependence
on the corresponding amplitudes.  For the fundamental mode this dependence is
almost exactly linear.  The plots of the $\varphi_{21}$ Fourier phases are very
flat because there are no resonances.  The difference between the
$\varphi_{21}$ value of the DM solution and that of the SM limit-cycle
is less than 0.2. The $\varphi_{31}$ parameters have the same tendencies but
with a little bit stronger amplitude dependence for the F mode.

\begin{figure}[ht]
\centerline{\psfig{figure=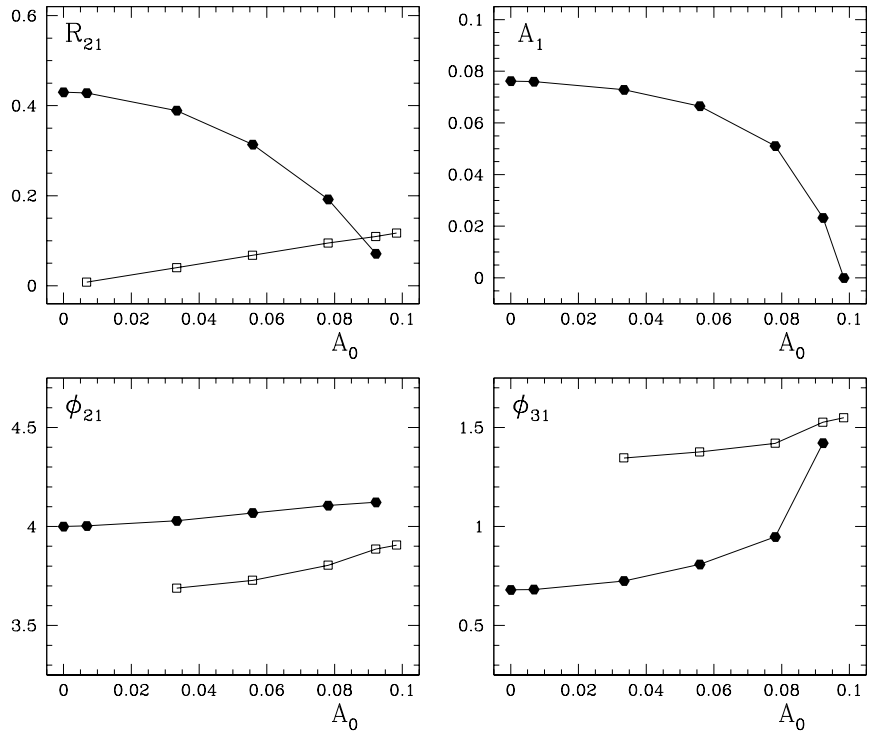,width=11.5cm}}
\caption{ The dependence of the Fourier parameters on the 
fundamental mode 
amplitude along the separatrix. Filled hexagons: fundamental mode, open 
squares: first overtone. The top right box shows the models on 
the ($A_0,A_1$) phase-space.}
\label{figfp}
\end{figure}

\subsection*{The Period ratios}

The period ratios offer an additional constraint for 
determining the physical parameters
of DM Cepheids and RR Lyrae stars (\eg Kov\'acs in this Volume). 
These investigations make use of the linear periods and the latter 
are usually based on radiative pulsation models.  Thus it is
very important to check how turbulent convection and nonlinear effects
affect the period ratios.

\vskip 10pt
{\bf Turbulent convection}
\vskip 5pt

The dominant source of period shifts between the radiative and turbulent
convective models is the convection induced change in the structure.  In the
radiative models there is a huge temperature gradient around the hydrogen
partial ionization zone, because radiation is not sufficiently effective in
transporting energy.  In the turbulent convective models the convective flux
carries some energy and thus reduces the temperature gradient.  This change in
the stellar structure causes a decrease in the period ratios.  The shift
depends on the efficiency of the convection, and thus is larger for lower
temperature models.  Since the convective flux parameter ($\alpha_c$) has not
been satisfactorily calibrated so far, it is not possible to state definitive
values for the shifts in the period ratios.  Depending on the parameters of the
models one finds $(P_1/P_0)_{tc}-(P_1/P_0)_{rad}\approx$ --0.0005 to --0.002.
Turbulent viscosity causes an additional shift of the periods, but that is
small compared to the effect of the convective flux.

\vskip 10pt
{\bf Nonlinearity}
\vskip 5pt

The magnitude of the nonlinear period shifts has remained unknown due to the
lack of nonlinear DM pulsation models.  In order to isolate the period shifts
from the larger period variations with \Teff\ we have used the time-frequency
analysis of the model output together with the amplitude-equation formalism,
instead of checking only the periods in the stationary DM oscillations.

The phase that is calculated from the analytic signal (Eq.~\ref{eqaf}) is
sufficiently smooth to yield the instantaneous frequencies by a numerical time
derivative of $\varphi(t)$.  Hydrodynamical integrations with a set of
different initial perturbations then can be used to fit the coefficients in the
phase equation (Eq.~\ref{phes}).  Since the nonlinear correction of the
frequencies are small compared to $\omega_j$, the period ratios can be fitted
directly with monomials
\begin{equation}
{ (P_1/P_0)_{nl} - (P_1/P_0)_{lin} \over  (P_1/P_0)_{lin} } 
= \sum C_{k,l} \th a_0^k a_1^l .
\label{eqpf}
\end{equation}

Fig.~\ref{figprs1} and \ref{figprs2} shows examples of the hydrodynamical
tracks that have been used to extract the nonlinear instantaneous period shift
as a function of the amplitudes in the transient states.  Fig~\ref{figprs1}
refers to an RR~Lyrae model (M=0.77\Mo, L=50\Lo) and Fig~\ref{figprs2} to a
Cepheid model (M=4.0\Mo, L=1100\Lo).  Although this fit gives an interpolation
formula for a specific model (with given mass, luminosity, and effective
temperature) only, the fit is found to vary only weakly inside the DM region.

\begin{figure}[ht]
\centerline{\psfig{figure=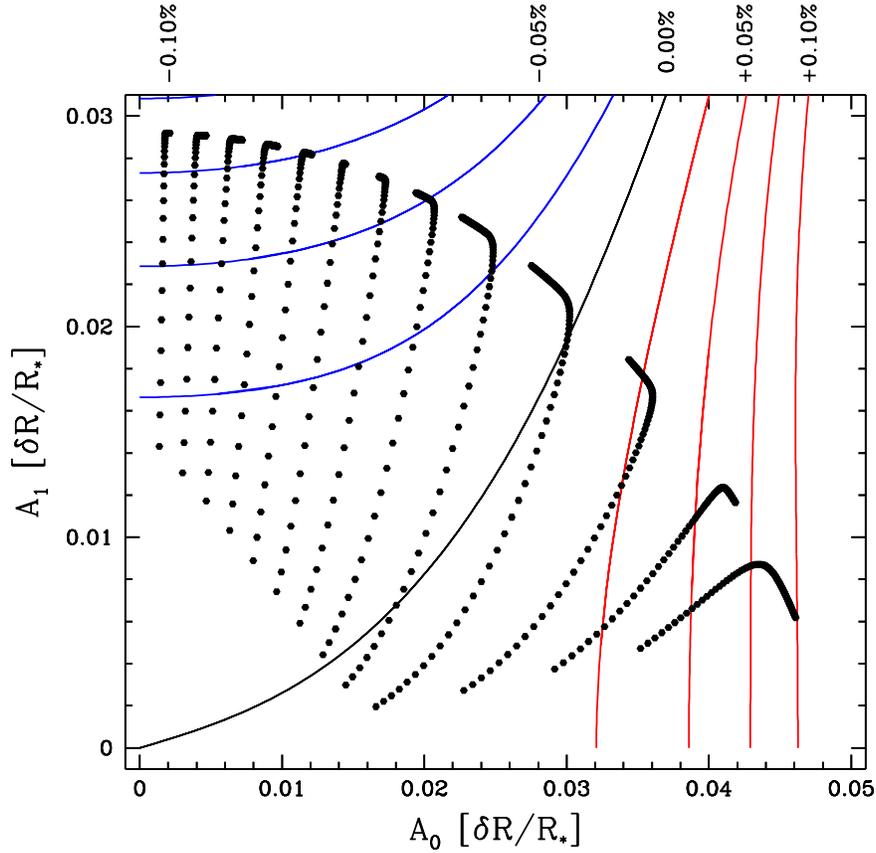,width=11.5cm}}
\caption{Nonlinear period shift of an RR Lyrae 
model in the ($A_0,A_1$) phase-space. }
\label{figprs1}
\end{figure}

\begin{figure}[ht]
\centerline{\psfig{figure=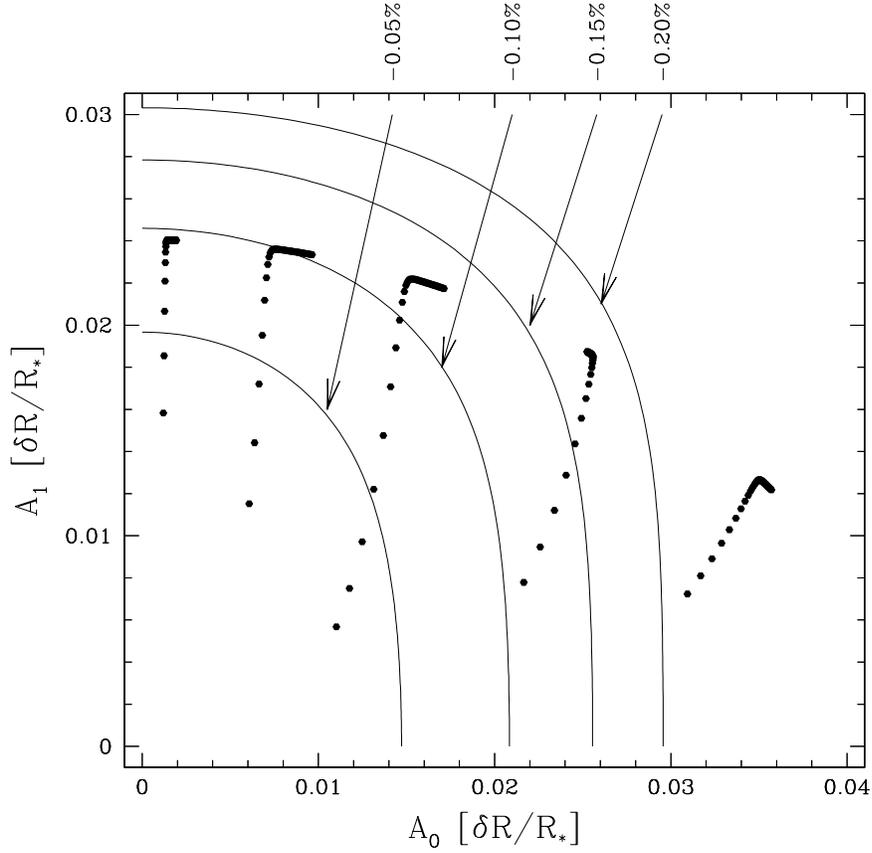,width=11.5cm}}
\caption{Nonlinear period shift of a
Cepheid model in the ($A_0,A_1$) phase-space. }
\label{figprs2}
\end{figure}

The 'equi-period-ratio' curves in the phase-space of the amplitudes are also
shown in Fig~\ref{figprs1} and \ref{figprs2}.  The relative changes of the
nonlinear (TC) period ratios compared to the linear ones are indicated in the
figures.  For the Cepheid model the period ratio is always smaller than the
linear one.  In contrast, for RR~Lyrae models the shift can be both positive
and negative.  These trends in the nonlinear shifts of the period-ratios have
been found to persist in our model calculations throughout a wide range of
model parameters.  Since the vast majority of the RR Lyrae stars pulsate with
higher amplitude in the first overtone than the fundamental mode, the
observed period ratios are likely to be smaller than the linear values.

Combining the two results we can conclude that the nonlinear turbulent
convective period ratios are thus generally, but not always, smaller than the
linear radiative ones, and that the relative difference is of the order of
several tenths of a percent.

\subsection*{Modal selection -- Bifurcation Diagram}

The model calculations with different initial perturbations of the static model
give the possible fixed point solutions of the star.  The behavior of the
individual models can be classified into five groups:\\
 (1) first overtone only (O),\\
 (2) fundamental mode only (F),\\
 (3) either fundamental or first overtone (E),\\
 (4) DM only (D), and \\
 (5) either fundamental or DM (H).\\
 We have not encountered any model where the first overtone and the DM fixed
points are simultaneously stable, but in principle such a situation is
possible.  For models with `E' or `H' characteristics there is hysteresis and
one should examine a sequence along the possible evolutionary path of the star
to predict its pulsation state.  Sequences with different temperatures, but
with a fixed luminosity, give good estimates for that purpose.

\begin{figure}[ht]
\centerline{\psfig{figure=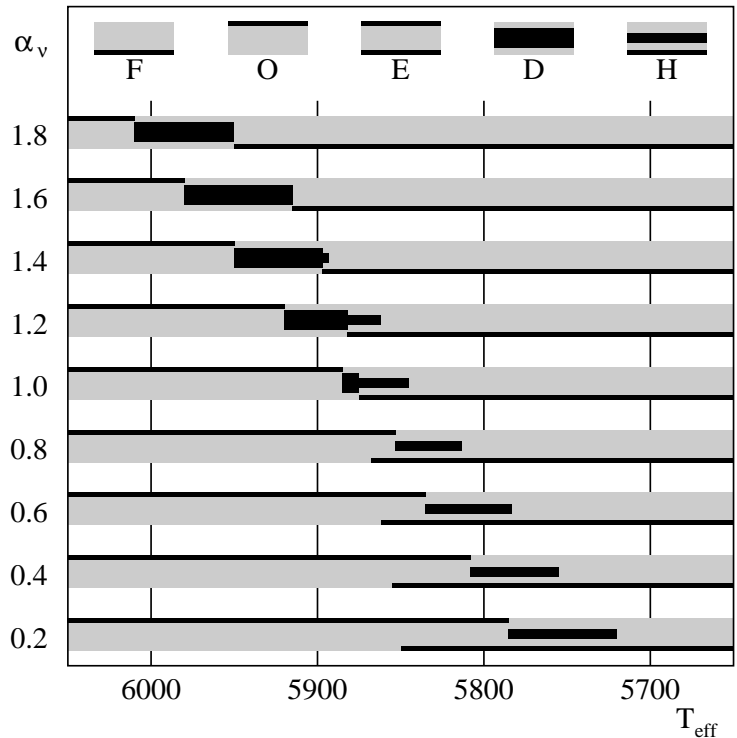,width=10cm}}
\caption{ Modal selection on the ($T_{\rm eff},\alpha_v$) plane for a Cepheid
model.  
The symbols of the modal states are described at the top of the figure. }
\label{fignu}
\end{figure}

In Fig.~\ref{fignu} the dependence of the modal selection on the eddy viscosity
is shown for a Cepheid model sequence with SMC composition.  The Galactic
Cepheids and the RR~Lyrae models show very similar behavior (for similar
$\alpha$ parameters).  The consistency of this similarity is somewhat
astonishing in light of the serious discrepancies that one finds between the
the observations and the calculations of low metallicity single-mode Cepheids
(to be discussed elsewhere).

\begin{figure}[ht]
\centerline{\psfig{figure=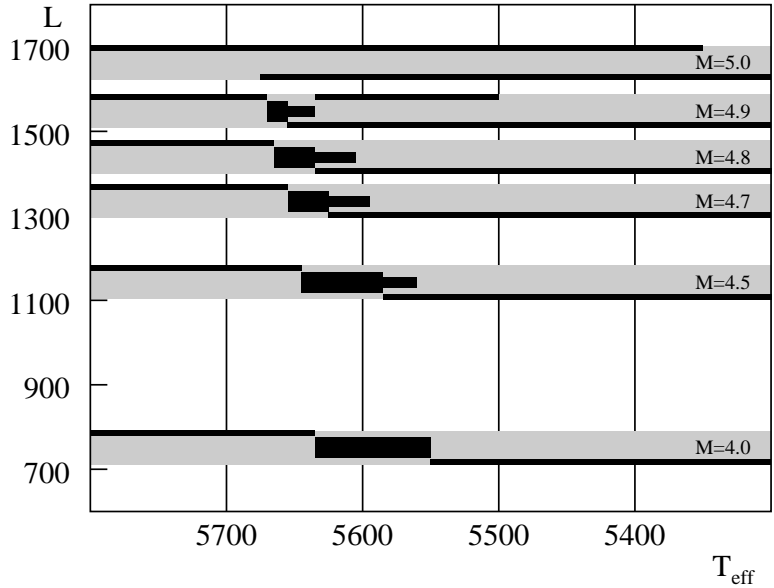,width=10cm}}
\caption{ Modal selection of galactic Cepheid model sequences
on the HR diagram. See Fig.~\ref{fignu} for the notation}
\label{fighrd1}
\end{figure}

\begin{figure}[ht]
\centerline{\psfig{figure=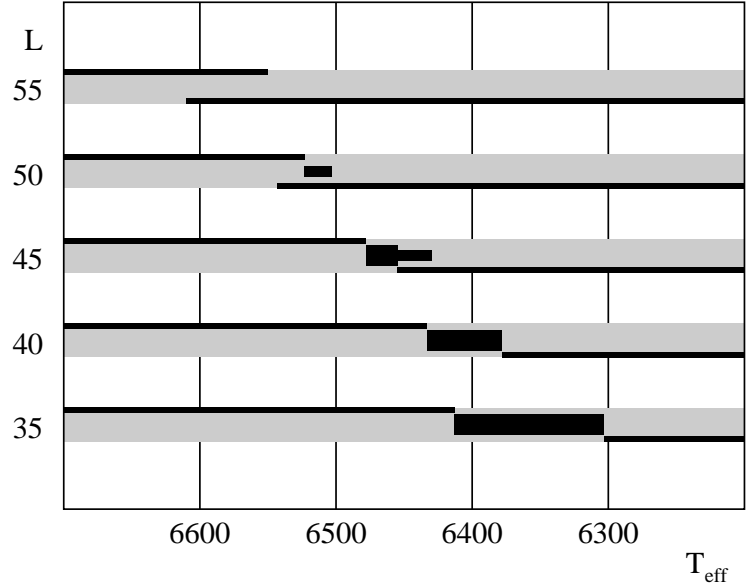,width=10cm}}
\caption{ Modal selection of galactic Cepheid and RR Lyrae model sequences
in the HR diagram. See Fig.~\ref{fignu} for the notation.}
\label{fighrd2}
\end{figure}

Turbulent viscosity has two dominant effects.  The first one is to shift the
transition region to higher temperatures (and to shorter periods) at higher
$\alpha_\nu$ values.  The second one is to changes the characteristics of modal
selection.  For high viscosity parameters a DM-only solution exists sandwiched
between the O and F state \th (we denote this transition by O-D-F).  As the
eddy viscosity decreased an 'either F or DM' (H) state appears at temperatures
between the D and F solutions (O-D-H-F transition).  For low $\alpha_\nu$
values the characteristics of the transition changes to O-E-H-F \ie the DM only
region is changed to an 'either-or' one.  

For different stellar parameters (see Fig.~\ref{fighrd1})  in
addition to the transition types seen in Fig.~\ref{fignu} we could identify the
following variations of modal behavior: O-E-F and O-D-H-E-F sequence.

For the O-D-F type sequence DM behavior occurs independently of the 
direction of evolution. However, for the O-E-H-F case DM pulsation 
can be observed only if the star evolves from the blue  side of the 
instability strip to the red.  In that case the star evolves to the 
DM state, when the O$_1$ mode  loses its stability, while on the opposite 
evolutionary path the star moves  to the overtone directly from the 
F mode at the blue edge of the fundamental mode instability strip 
through a transient DM state.

The modal selection is illustrated in Figures~\ref{fighrd1} and \ref{fighrd2}
both for a Galactic Cepheid model and for an RR~Lyrae model in an HR diagram .

We have calculated the Cepheid models with a mass--luminosity relation (the
masses of the models are indicated at the right end of the gray stripes).  One
notes a very interesting feature, namely the disappearance of the DM solution
at higher luminosities, even where stable first overtone pulsation exists. This
result is in good agreement with the observations by Udalski \etal (1999).

Despite this very encouraging agreement we recall however that these
bifurcation diagrams remain tentative until a full calibration of the $\alpha$
parameters has been made.

\section{The Future}

Thanks to the incorporation of turbulent convection in the hydrodynamical codes
it is finally possible to model steady DM pulsations both in Cepheids and
RR Lyrae stars.  Some problems persist though.  The large-scale surveys have
found second overtone to first overtone (O$_2$/O$_1$) DM pulsators at similar
population as O$_1$/F beat Cepheids.  The models should reproduce this
constraint as well.  With the recent parameter setting we could find some
O$_2$/O$_1$ models, but in a very narrow range of temperature and turbulence
parameters.  To model these DM variables satisfactorily, we have to explore the
$\alpha$ parameter space in a wider range and perhaps improve the description
of turbulent convection. 

This effort is important because a consistent modelling of DM oscillations in
both Cepheids and RR Lyrae stars and with different chemical compositions is
necessary for understanding the physical mechanisms inside the stars, on the
one hand, and for enhancing our confidence that we can use these stars reliably
as standard candles, on the other hand.

\begin{acknowledgments}

This work has been supported by the Hungarian OTKA (T-026031) grant and by the
National Science Foundation (AST9819608).  It is a great pleasure to
acknowledge the collaboration of Jean-Philippe Beaulieu, Zolt\'an Csubry,
R\'obert Szab\'o, and Phil Yecko on this project.  We wish to thank
Michael Feuchtinger for a thorough reading of the manuscript.

\end{acknowledgments}

\end{document}